\documentclass[prl,aps,amsmath,amssymb,showpacs,floatfix,twocolumn]{revtex4-1}
\usepackage{graphicx,epsfig}
\usepackage[usenames]{color}
\usepackage{bbold}

\newcommand{\beq}{\begin{equation}}
\newcommand{\eeq}{\end{equation}}
\newcommand{\bea}{\begin{eqnarray}}
\newcommand{\eea}{\end{eqnarray}}
\newcommand{\marton}[1]{#1}
\newcommand{\martonnew}[1]{#1}
\newcommand{\martonnewnew}[1]{#1}
\newcommand{\leonid}[1]{#1}

\begin{document}
\title{Anomalous conductances in an ultracold quantum wire}
\author{M. Kan\'asz-Nagy$^{1}$, L. Glazman$^{2}$, T. Esslinger$^{3}$, and E. A. Demler$^1$}
\affiliation{$^1$Department of Physics, Harvard University, Cambridge, MA 02138, U.S.A.}
\affiliation{$^2$Department of Physics, Yale University, New Haven, CT 06520, U.S.A.}
\affiliation{$^3$Department of Physics, ETH Zurich, 8093 Zurich, Switzerland}

\begin{abstract}
We analyze the recently measured anomalous transport properties of an ultracold gas through a ballistic constriction [S. Krinner {\it et al.}, PNAS 201601812 (2016)]. The quantized conductance observed at weak interactions increases several-fold as the gas is made strongly interacting, which cannot be explained by the Landauer theory of single-channel transport. We show that this phenomenon is due to the multichannel Andreev reflections at the edges of the constriction, where the interaction and confinement result in a superconducting state. Andreev processes convert atoms of otherwise reflecting channels into the condensate propagating through the constriction, leading to a significant excess conductance. Furthermore, we find the spin conductance being suppressed by superconductivity; the agreement with experiment provides an additional support for our model.
\end{abstract}

\maketitle

\begin{figure}
\includegraphics[width=8.0cm,clip=true]{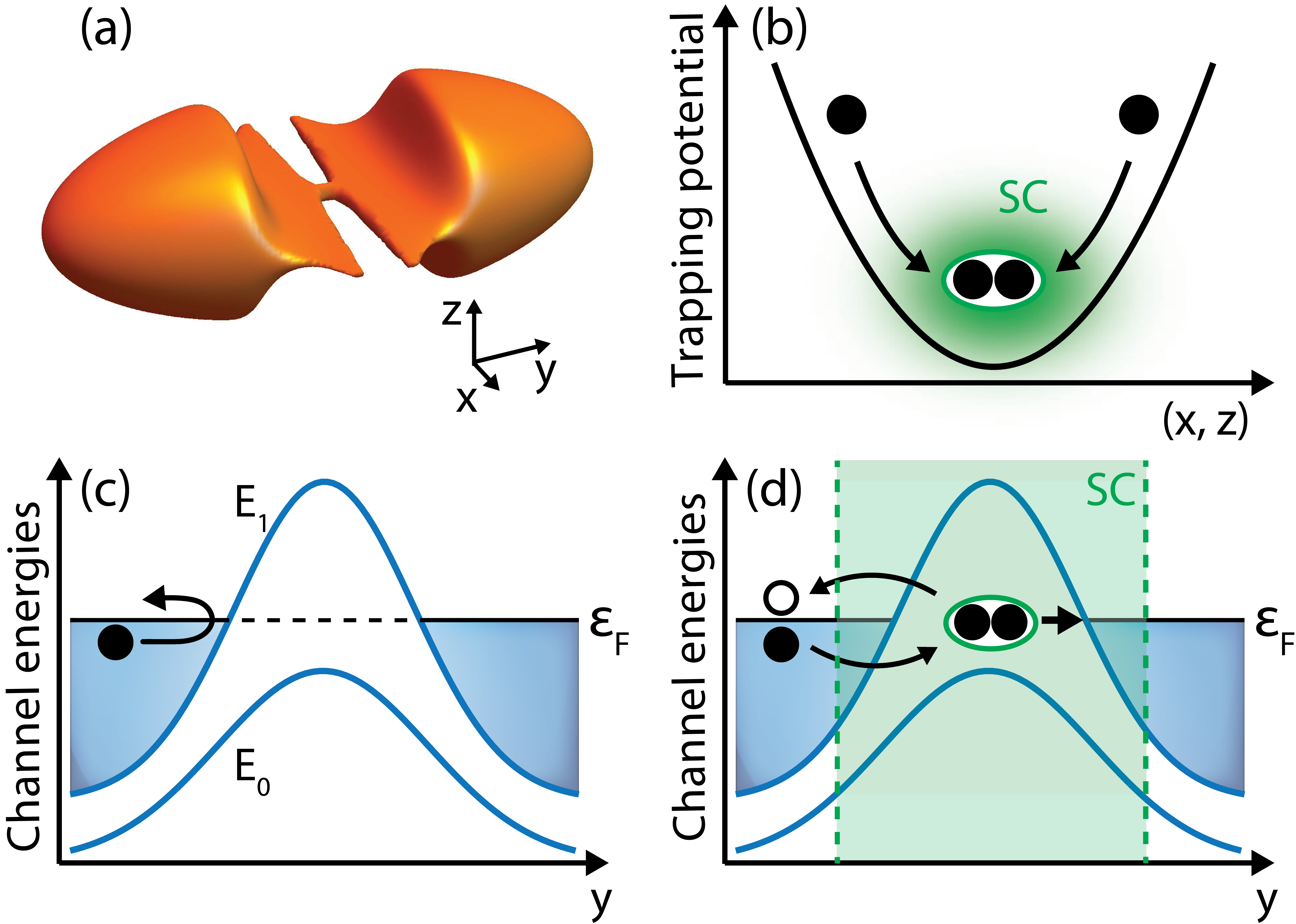}
\caption{(Color online.) 
(a) Geometry of the ultracold gas. The center of a trap is optically confined into a one-dimensional constriction, surrounded by a two-dimensional region, connected to three-dimensional reservoirs. 
(b) SC pairing is possible between channels of arbitrarily high transverse \marton{modes} due to their non-zero coupling to the condensate.
(c) Transport through the one-dimensional constriction at weak interactions: only the lowest channel is transmitting, providing $1/h$ conductance.
(d) Pairing at strong interactions lead to Andreev processes at the SC-normal interface in higher channels as well, contributing significantly to the conductance.
}
\label{fig:setup}
\end{figure}

Transport measurements through one-dimensional ballistic channels provide invaluable insight into the complex many-body systems by connecting microscopic quantum  dynamics with macroscopic observables, such as the conductance $G_n$, spin conductance $G_s$ and heat transport. In the \marton{normal state}, these quantities exhibit plateaus as a function of the gate potential at integer multiples of the conductance and heat conductance quantum, respectively~\cite{cond_quant_measurement,spin_cond_quant_measurement,heat_cond_quant_measurement}. If the channel or leads are made superconducting, a wealth of other phenomena opens up. At a normal-superconducting interface, a fermion incident from the  normal metal to the superconductor forms a Cooper pair with another fermion so that they can enter the condensate, \marton{while} a hole gets reflected from the interface -- a process called  Andreev reflection (AR)~\cite{Andreev_reflection, Andreev_and_proximity_1, Andreev_and_proximity_2, Andreev_and_proximity_3}. AR lies at the heart of several interesting transport phenomena, including Andreev bound states~\cite{Andreev_and_proximity_2}, Shiba states~\cite{Shiba_states}, \martonnew{manifestation of the} charge parity effect in superconducting grains~\marton{\cite{even_odd_SC,even_odd_SC_Leonid,Tinkham_NSN}}, quantum Andreev oscillations~\cite{QAndreev_oscillations}, superconducting spintronics~\cite{SC_spintronics}, Cooper pair splitting~\cite{Cooper_pair_splitter}, as well as the celebrated Majorana states of topological superconductors~\cite{Majorana_th, Majorana_exp_1, Majorana_exp_2, Majorana_exp_3, Majorana_exp_4}.
Despite the abundance of exotic transport phenomena in electronic condensed matter systems, it has been only very recently that the conductance properties of charge neutral massive particles have been measured, using an ultracold Fermi gas of $^6{\rm Li}$ atoms, passed through an optically created one-dimensional constriction, realizing the limiting case of a ballistic wire of a single transmitting transverse channel~\cite{Esslinger_weak_int, Esslinger_nonlinear_cond, Esslinger_strong_int}, \marton{see Fig.~\ref{fig:setup}. This system offers tunability of the geometry and interactions, with the opportunity to reach the strongly interacting regime, where the wire becomes superconducting, contacted by normal leads in the experiment of Ref.~\onlinecite{Esslinger_strong_int}. In condensed matter environments, similar systems of inhomogenous superconductivity have attracted significant attention, \martonnew{providing access to phenomena} on the verge between microscopic and mesoscopic physics, such as phase-slips~\cite{Tinkham_phase_slips}, non-local quantum correlations~\cite{Chandrasekhar} and spatially resolved AR~\cite{Klapwijk}. Superconducting islands immersed in a metallic environment \martonnew{may also comprise a platform for the study of} the superconductor-metal \martonnew{transition}~\cite{SC_metal_1,SC_metal_2,SC_metal_3,SC_metal_4,SC_metal_5,SC_metal_6,SC_metal_7}.}

In the presence of weak interactions, the constriction exhibits conductance plateaus of integer multiples of the $1/h$ as a function of the confinement strength, in accordance with several similar experiments in ballistic nanostructures~\cite{cond_quant_measurement,spin_cond_quant_measurement,Esslinger_weak_int,Landauer_theory}. 
Rather surprisingly, however, making the gas strongly interacting leads to larger than four-fold increase in the conductance of the lowest plateau of a single transverse \marton{mode}. This is in apparent contradiction with the simple Blonder--Tinkham--Klapwijk (BTK) model of transport through a single ballistic channel~\cite{Landauer_theory,BTK_paper}: although interactions can make the channel superconducting (SC), this can at most lead to a factor of $2$ increase in the conductance, \marton{since in AR each incident atom drags along at most another atom through the constriction, as a Cooper pair.}

\marton{We resolve the puzzle of anomalous conductance by associating it with multichannel AR processes at the normal-superconductor \martonnew{interfaces at the two ends} of the constriction (see Fig.~\ref{fig:setup}). Confinement significantly renormalizes the interactions within the central part of the constriction, leading to strong SC pairing~\cite{Parmenther_NSN}. This pairing field penetrates into the normal leads, with several channels below the Fermi energy. Atoms in higher transverse modes, that would otherwise be reflected by the constriction, can go through AR processes within this thin superconducting interface. As they become part of the condensate they propagate through the junction \martonnew{as Cooper pairs}~\cite{Andreev_and_proximity_2, proximity_effect}\martonnew{;} the resistance of the channel is entirely determined by the interface~\cite{resistance_at_SC_normal}. Furthermore, as the interaction increases, current is increasingly carried by Cooper pairs, the spin current approaches zero. \martonnew{This agrees with the experimental observations} of Ref.~\onlinecite{Esslinger_strong_int}.}

\begin{figure}[t]
\includegraphics[width=8.5cm,clip=true]{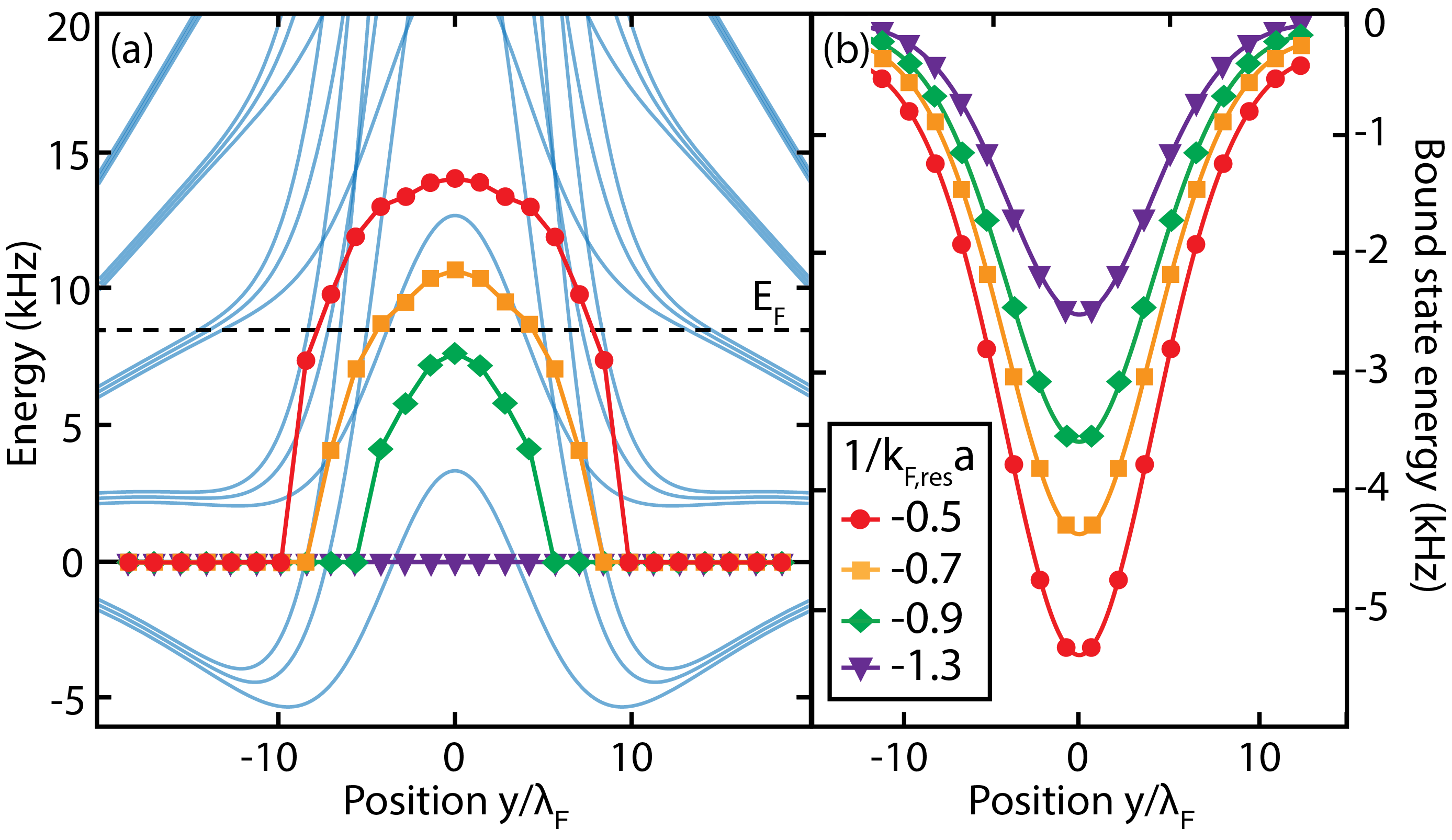}
\caption{(Color online.) 
(a) SC pairing amplitude $\Delta_0$ (lines with symbols) across the constriction at different interaction strengths shown in the inset of (b). Solid lines indicate the energies of the transverse modes of different $n_z$ quantum numbers, with only $n_x = 0, 1$ and $2$ \marton{modes} shown. At the edges of the constriction, the gas is quasi-two dimensional, and the $n_x$ \marton{modes} are almost completely degenerate. They split near the middle, where the gas becomes quasi-one dimensional.
(b) Bound state energies along the constriction, renormalized by the confinement.
[Parameters: $\nu_x(y), \nu_z(y)$ and $V_g(y)$ are approximated as Gaussians of
HWHM $(d_x, d_y, d_V) = (4.7, 17.7, 15.1)~{\rm \mu m}$, as in Ref.~\onlinecite{Esslinger_strong_int},
and heights $(\nu_{x0},\nu_{z0}, V_{g0}) = (23.2~{\rm kHz}, 9.2~{\rm kHz}, 0.625~{\rm \mu K})$. Cut-off in the channel number: $n_x, n_z \leq 8$. $\lambda_F$ and $E_F$ denote the Fermi wavelength and the Fermi energy within the reservoirs.]
}
\label{fig:energies}
\end{figure}

The experimental geometry is shown in Fig.~\ref{fig:setup}~(a). The central part of the gas is squeezed into two dimensions using lithographic imprinting, whereas a narrower perpendicular laser beam pinches the middle of this region to form a short one-dimensional ballistic quantum wire~\cite{Esslinger_weak_int,Esslinger_nonlinear_cond,Esslinger_strong_int}. The conductance of the wire is tunable either by tuning the confinement frequencies $\nu_{x0}, \nu_{z0}$, or using a gate potential $V_{g0}$, created by an additional, wide laser beam along the $z$ axis (see the caption of Fig.~\ref{fig:energies}).
By creating a density or spin imbalance between the two sides of the junction, the conductance $G_n$ and spin conductance $G_s$ can be determined by monitoring the relaxation of the population imbalance in time, and making use of the equation of states of the gas within the leads~\cite{Esslinger_weak_int, Esslinger_nonlinear_cond}.

We determine the superconducting profile in the constriction within the local density approximation (LDA), whereby we consider a small part of the system of length $L_y$, where the parameters of the gas are assumed to be constant. We also take into account the renormalization of interactions due to confinement effects. The constriction is described by a harmonic Hamiltonian of trapping frequencies $\boldsymbol{\omega} = (\omega_x, \omega_z) = (\nu_x, \nu_z)/2\pi$, local gate potential $V_g$ and chemical potential $\mu$,
\beq
H_{\rm kin} = \sum_{{\bf n},\sigma, q} \xi_{{\bf n}, \sigma, q} \, a^\dagger_{{\bf n}, \sigma, q} a_{{\bf n}, \sigma, q},
\eeq
where $\xi_{{\bf n}, q} = \frac{\hbar^2 q^2}{2 m} - V_g -\mu_\sigma + (n_x + \frac{1}{2})\hbar\omega_x + (n_z + \frac{1}{2})\hbar\omega_z$ denotes the channel energies, and $a_{{\bf n}, q, \sigma}$ annihilates an atom in channel ${\bf n}=(n_x,  n_z)$, spin $\sigma = \uparrow, \downarrow$ and momentum $q$ along the $y$ axis. The interaction between the $^6{\rm Li}$ atoms is given by the standard point interaction $g \, \delta^{(3)}({\bf r})$, where $g$ is the bare interaction strength~\cite{Bloch_review}. In order to simplify the treatment of the interaction term, it is worth going into the center of mass (COM) and relative frame of the colliding atoms along the trapped directions, $(x_1,x_2) \to \left( \frac{x_1+x_2}{2}, x_1-x_2 \right)$, and similary for $z$, with the coordinates of the atoms denoted by $(x_1, z_1)$ and $(x_2, z_2)$. One can thus transform the interaction Hamiltonian according to the unitary transformation $\langle {\bf N}, {\boldsymbol \nu}| {\bf n}_1, {\bf n}_2 \rangle$, where ${\bf N} = (N_x, N_z)$ and ${\boldsymbol \nu} = (\nu_x, \nu_z)$ denote the COM and relative harmonic oscillator quantum states, and ${\bf n}_1$, ${\bf n}_2$ stand for those in the laboratory frame. These matrix elements are non-zero only for ${\bf n}_1 + {\bf n}_2 = {\bf N} + {\boldsymbol \nu}$ combinations, due to energy conservation. Since harmonic trapping and interactions both conserve ${\bf N}$ and the COM momentum $Q$, the interaction Hamiltonian can be decoupled exactly as~\cite{2D_BCS}
\bea
H_{\rm int} &= & \frac{1}{\tilde g} \; \sum_{{\bf N}, Q} \hat{\Delta}^\dagger_{{\bf N}, Q} \hat{\Delta}_{{\bf N}, Q}, \label{eq:H_int}\\
\hat{\Delta}_{{\bf N}, Q} &\equiv & \tilde{g} \sum_{{\bf n}_1, {\bf n}_2, k} V_{\bf N}^{{\bf n}_1 {\bf n}_2} 
a_{{\bf n}_1, \downarrow, Q-k}^\dagger a_{{\bf n}_2, \uparrow, k},
\eea
where the interaction strength $\tilde{g} = g/(l_x L_y l_z)$ of energy dimension is defined using oscillator lengths $l_{x(z)} = \sqrt{\hbar/m\omega_{x(z)}}$. The matrix elements $V_{\bf N}^{{\bf n}_1 {\bf n}_2} = \varphi_{\nu_x}(0) \, \varphi_{\nu_z}(0) \langle {\bf N}, {\boldsymbol \nu}| {\bf n}_1, {\bf n}_2 \rangle$, with ${\boldsymbol \nu} = {\bf n}_1 + {\bf n}_2 - {\bf N}$, arise from the matrix elements of the point-like interaction potential~\cite{2D_BCS} (see Supplementary Material). In the previous expression, the value of the relative harmonic oscillator wave function $\varphi_\nu$ at the origin is given by $\varphi_\nu(0) = \frac{(-1)^{\nu/2}}{(2\pi)^{1/4}} \sqrt{\frac{1}{2^\nu} \binom{\nu}{\nu/2}}$ for $\nu$ even and $\varphi_\nu(0) = 0$ for $\nu$ odd.

We decouple Eq.~\eqref{eq:H_int} in a standard BCS approximation $\Delta_{\bf N} \equiv \langle \hat{\Delta}_{{\bf N}, Q=0} \rangle$. Although in general it could be possible to have SC ordering in many COM \marton{modes}, we verified that for the experimental parameters of Ref.~\onlinecite{Esslinger_strong_int} considered here, only the ${\bf N}=0$ \marton{mode} gains non-zero pairing amplitude. Thus, in the following, we focus on this case and leave the general discussion to the Supplementary Material. The resulting Bogoliubov-de Gennes Hamiltonian for quasi-particle excitations reads
\beq
H_{BdG} = \sum_{q} \left( {\bf a}^\dagger_{\uparrow, q}, {\bf a}_{\downarrow, -q} \right)
\begin{pmatrix}
\boldsymbol{\xi}_{\uparrow,q} & \boldsymbol{\Delta}^\dagger \\
\boldsymbol{\Delta} & -\boldsymbol{\xi}_{\downarrow,q}
\end{pmatrix}
\begin{pmatrix}
{\bf a}_{\uparrow, q} \\
{\bf a}^\dagger_{\downarrow, -q}
\end{pmatrix},
\label{eq:H_BdG}
\eeq
in vectorial notation for the band indices. Here, $\left({\bf a}_{\sigma, q}\right)_{\bf n} = a_{{\bf n},\sigma,q}$ denotes the vector of annihilation operators, the SC matrix is given by $\boldsymbol{\Delta}_{{\bf n}_1{\bf n}_2} = \Delta_{0} V_{0}^{{\bf n}_1{\bf n}_2}$, and 
the matrix $\boldsymbol{\xi}_{\sigma,q}$ contains the band energies on its diagonal.
Using a Bogoliubov transformation, one can now determine the quasi-particle energies $E_{{\bf n},q}$. Then, in order to determine the pairing amplitude $\Delta_0$, one needs to minimize the finite temperature BCS free energy 
$F_{MF} = E_{MF} - T \sum_{{\bf n}, q} \log\left( 1 + e^{-E_{{\bf n},q}/T} \right)$
at a fixed chemical potential, as set by the leads.
The mean-field condensation energy $E_{MF} = \sum_{{\bf n}, q} \left(\xi_{{\bf n}, q} - E_{{\bf n}, q} \right) - \frac{|\Delta_0|^2}{\tilde g}$ however still contains the bare interaction term $\tilde g$, and a divergent sum over excitation energies. In order to regularize this term, we make use of the vacuum Bethe--Salpeter equations~\cite{CIR_pietila,CIR_kanasz}, and express $g$ in terms of physical quantities: the scattering length $a$ or, equivalently, the vacuum bound state energy $E_B$ (see Supplementary Material),
\bea
\frac{1}{g} &=& \frac{m}{4\pi \hbar^2 a} - \int\frac{d^3 q}{(2\pi)^3} \frac{m}{\hbar^2 q^2 + i 0^+}  \label{eq:g_reg} \\
&=& \frac{1}{l_x l_z} \int \frac{dq}{2\pi} \sum_{{\bf n}_1 {\bf n}_2} \frac{\left| V_0^{{\bf n}_1 {\bf n}_2} \right|^2}{E_B - \left( \frac{\hbar^2 q^2}{m} + \hbar ({\bf n}_1 + {\bf n}_2) \boldsymbol{\omega} \right)}. \notag
\eea
In contrast to three-dimensional systems, Eq.~\eqref{eq:g_reg} always has a bound state solution $E_B<0$ in quasi-one dimensional gases, even on the attractive side of the Feshbach resonance~\cite{Bloch_review,CIR_Olshanii,Tonks_Girardeau}. As we show in Fig.~\ref{fig:energies}~(b), $E_B$ strongly depends on the confining frequencies as well as on the scattering length, and incorporates the confinement-induced renormalization of the interaction.
Making use of Eq.~\eqref{eq:g_reg}, we can now express the condensation energy in terms of $E_B$, and, as we show in the Supplementary Material, the resulting expression is regular,
\beq
E_{MF} = \sum_{{\bf n}, q} \left(\xi_{{\bf n}, q} - E_{{\bf n}, q} \right)
- \sum_{{\bf n}_1, {\bf n}_2, q} \frac{|\Delta_0|^2 \, \left| V_0^{{\bf n}_1 {\bf n}_2} \right|^2}{E_B - \left( \frac{\hbar^2 q^2}{m} + \hbar ({\bf n}_1 + {\bf n}_2) \boldsymbol{\omega} \right)}. \label{eq:E_MF}
\eeq

\begin{figure}[t]
\includegraphics[width=8.5cm,clip=true]{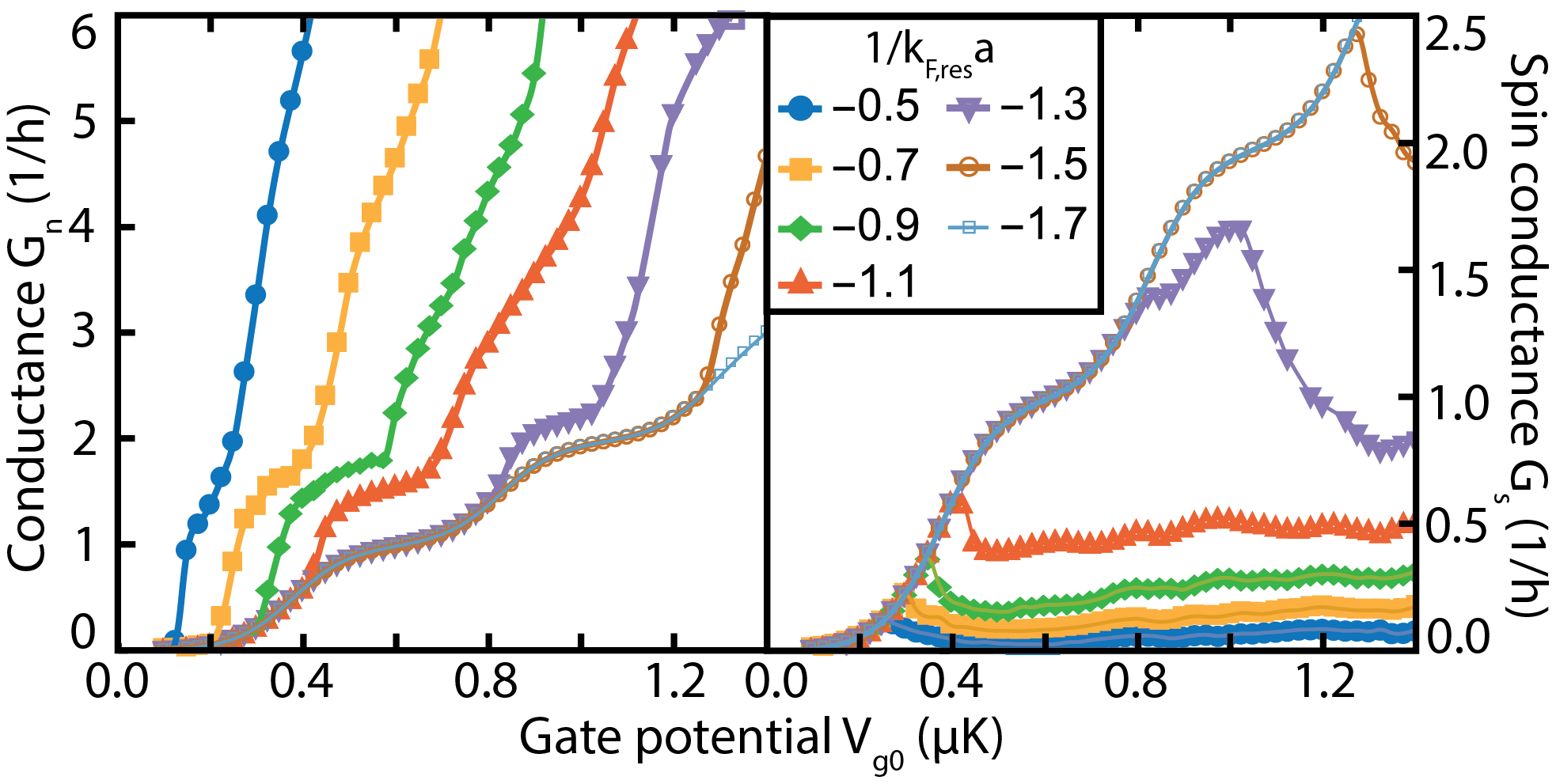}
\caption{(Color online.) 
Conductance (a) and spin conductance (b) as a function of the gate potential at at different interaction strengths $1/(k_{F,res} \, a)$ in the reservoirs. 
[Parameters: $T=62~{\rm nK}$, $\mu = 8.5 ~{\rm kHz}$, $(\nu_{x0}, \nu_{z0}) = (9.2, 23.2)~{\rm kHz}$, the geometry is identical to the one in Fig.~\ref{fig:energies}.]
}
\label{fig:scan_Vg}
\end{figure}

Fig.~\ref{fig:energies}~(a) shows typical profiles of the SC order parameter $\Delta_0(y)$ at various interaction strengths. Due to strong confinement towards the middle of the constriction, the bound state becomes significantly deeper in energy favoring superconductivity in Eq.~\eqref{eq:E_MF}. Although in the middle there is only one channel below the Fermi energy that can contribute to pairing, higher \marton{transverse modes} are also coupled to the condensate \marton{in the SC-normal interface} through Eq.~\eqref{eq:H_BdG}. At the largest interaction strengths, the SC gap becomes comparable to the Fermi energy~\cite{SC_larger_than_EF}. This strong pairing also extends around the central potential hill of the constriction, providing a thin superconducting layer that is responsible for the excess conductance seen in the experiment~\cite{Esslinger_strong_int}, due to multichannel Andreev processes. 
The length scale over which these processes happen are of the order of the SC healing length $\tilde{\xi}_s$. Even though the width of this region is just a few times the Fermi wavelength $\lambda_F$, the strong pairing within the constriction leads to $\tilde{\xi}_s \sim \lambda_F$, and the AR probabilities become non-negligible.

\begin{figure}[t]
\includegraphics[width=8.5cm,clip=true]{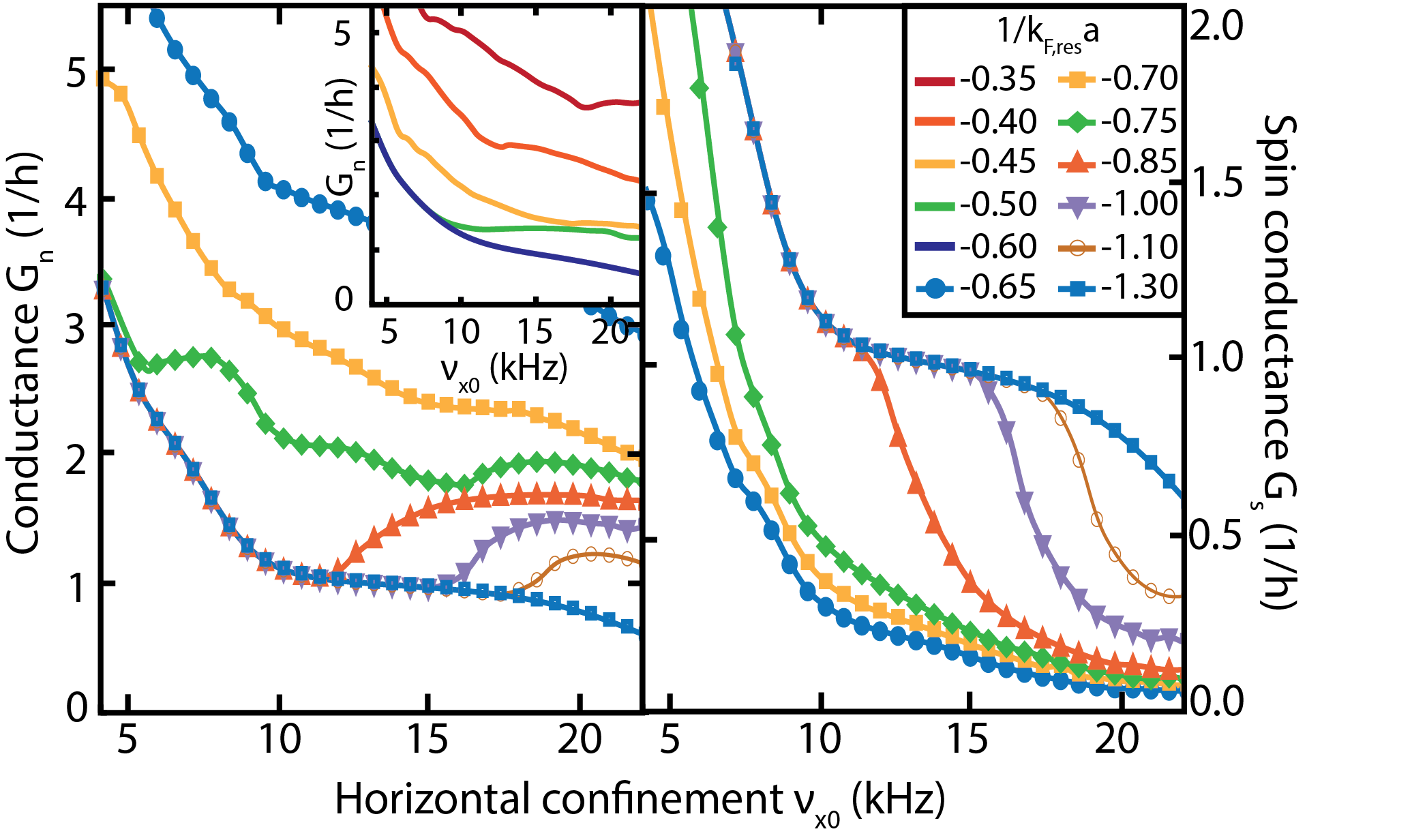}
\caption{(Color online.) 
Conductance (a) and spin conductance (b) as a function of the horizontal confinement at at different interaction strengths and at a temperature $T=62~{\rm nK}$. 
\marton{The conductance exhibits non-monotonic behavior due to the onset of SC at large confinement strengths, an effect that goes away at higher temperatures $T=109~{\rm nK}$, shown in the inset.}
[Parameters: $\mu = 8.5 ~{\rm kHz}$, $(V_{g0}, \nu_{z0},) = (100~{\rm nK}, 23.2~{\rm kHz})$, the geometry is identical to the one in Fig.~\ref{fig:energies}.]
}
\label{fig:scan_nux}
\end{figure}

We determine the conductance and spin conductance of the waveguide in a Landauer picture~\cite{Landauer_theory}, by calculating the reflection and AR coefficients $\left({\bf r}_{pp}\right)_{{\bf n}^\prime {\bf n}}$ and $\left({\bf r}_{hp}\right)_{{\bf n}^\prime {\bf n}}$, respectively, describing reflections from channel ${\bf n}$ to ${\bf n}^\prime$, with the $p$ and $h$ indices denoting particle and hole states. To do this, we  determine the eigenmodes of the Bogoliubov--de Gennes Hamiltonian Eq.~\eqref{eq:H_BdG} at all incoming energies $\epsilon$, (see Supplementary Material).
The zero bias conductance and spin conductance are given by a thermal average over these contributions~\cite{Landauer_theory},
\bea
G_{n/s} =- \int d\epsilon \, n_F^\prime(\epsilon-\mu) \, {\rm Tr}\left( \mathbb{1} - {\bf r}_{pp}^\dagger {\bf r}_{pp} \pm {\bf r}_{hp}^\dagger {\bf r}_{hp} \right), \label{eq:G}
\eea
where $\mathbb{1}$ denotes the unit matrix, $n_F$ stands for the Fermi function, and the energy arguments of ${\bf r}_{pp}(\epsilon)$ and ${\bf r}_{ph}(\epsilon)$ are neglected for brevity. As can be seen from Eq.~\eqref{eq:G}, AR processes contribute to the conductance, but they decrease the spin conductance. \martonnewnew{ The definition of the spin conductance in Eq.~\eqref{eq:G} differs from that of Ref.~\onlinecite{Esslinger_strong_int} by a factor of two, due to the ambiguity in defining the chemical potential difference in case of the spin current. Using the definition above, the spin and charge conductances are identical in the normal phase, and their deviation indicates the onset of superconductivity.}

As shown in Fig.~\ref{fig:scan_Vg}, both $G_s$ and $G_n$ show the usual Landauer quantization as a function of the gate potential $V_{g0}$ at weak interactions, as has been observed experimentally~\cite{Esslinger_weak_int}. At increasing interaction strengths, the constriction becomes superconducting, leading to increased conductance and suppressed spin conductance.  As $V_{g0}$ is tuned, SC order appears first in the middle of the constriction (see Fig.~\ref{fig:energies}), thus only the otherwise transmitting channels can participate in Andreev processes. This is the regime of the BTK theory, and we observe well defined plateaus, within a factor of two increase in conductance. \marton{At larger gate potentials, however, the number of channels in the superconducting interface increases, leading to a strong increase in conductance.} Since the SC layer at the end of the wire is thin, most channels cannot go through perfect ARs and they only contribute a small fraction of a conductance quantum to $G_n$. The plateaus thus become \marton{less well-defined}. Furthermore, in agreement with experiment~\cite{Esslinger_strong_int}, we find that $G_s$ depends non-monotonically on the gate potential in Fig.~\ref{fig:scan_Vg}~(b). The reason for this is that as $V_{g0}$ increases, additional channels are pulled down below the Fermi energy, and the system gains additional condensation energy by forming Cooper pairs in these channels. As a result, SC pairing increases, and a larger fraction of the current is carried by Cooper pairs, leading to a sudden drop in $G_s$.

Fig.~\ref{fig:scan_nux} shows $G_s$ and $G_n$ as a function of the horizontal confinement $\nu_{x0}$, exhibiting a broad conductance plateau at $(1/h)$ conductance at weak interactions. In agreement with the experiment~\cite{Esslinger_strong_int}, the conductance plateau is still somewhat visible at larger interaction strengths, but pushed to a much larger value due to superconductivity (see the curves $1/(k_{F,res} a) = -0.70$ and $-0.75$).
However, we also find an interesting non-monotonicity of the conductance curves at strong confinement, that has not been observed experimentally. This behavior is due to the confinement-induced renormalization of the interaction, that leads to the onset of SC at tighter confinements. This is  accompanied by a sudden decrease in the spin conductance (see Fig.~\ref{fig:scan_nux}~(b)). 
This non-monotonicity does not appear at higher temperature as the confinement-induced onset of pairing is killed by temperature fluctuations, see the inset of Fig.~\ref{fig:scan_nux}~(a). This effect thus may be observable by further cooling the gas in the experiment.

The comparison of Fig.~\ref{fig:scan_nux}~(a) and the inset also demonstrates the sensitivity of the conductance curves to experimental parameters, as also seen in Ref.~\onlinecite{Esslinger_nonlinear_cond}. As we show in the Supplementary Material, conductance at strong interactions, $1/(k_{F, res} a) \sim 0.5$, can change as large as a factor of 5 just by changing the temperature and chemical potential within their $\sim 15\, \%$ experimental error bars. The reason is that $\Delta_0$ depends very sensitively on these parameters near the onset of superconductivity, and its value has a significant influence on conductance. Further important uncertainties arise from experimental aberrations of the laser fields that form the constriction. Since the transport is largely governed by an interface effect at the edge of the \marton{constriction}, these geometric factors become important~\cite{Esslinger_nonlinear_cond}.

\marton{As an experimental test of our theory, we propose to investigate the channel's conductance at large, equal spin imbalances in both leads, leading to the suppression of the constriction's superconductivity due to Fermi surface mismatch. At large imbalances, the SC-normal transition could thus be measured using the drop of anomalous conductances, and from the increase of spin conductance, to their respective values in the normal state~\cite{Sarma,Clogston_limit,Ketterle_imb_SF}.}

\leonid{
The above analysis of quantum transport assumes a static order parameter in the superconducting region. Its finite size may constrain the fluctuations of the number of atoms in the region. The constrained particle number fluctuations enhances the fluctuations of phase of the order parameter. These effects were studied extensively in the context of Coulomb blockade in a superconducting island coupled to a normal-metal lead, see, e.g. Refs.~\onlinecite{leonid1}~and~\onlinecite{leonid2}. The overall conclusion is that at large conductance of the interface the effects of Coulomb blockade (i.e., constraints on the particle number) are negligible. The corresponding energy scale turns out to scale as exponent of $-G/G_q$ if the large conductance of a junction is achieved by increasing the number of conducting channels \cite{leonid1, leonid2}, and as a product of reflection amplitudes in each of the channels, in case of an arbitrary (even small) number of highly-transparent channels \cite{leonid2,leonid3}. The phase fluctuations are small, and their estimate in the Gaussian approximation is provided in Section 6 of the Supplementary.
}

{\it Conclusion --} We demonstrated that the recently observed anomalous transport measured in Ref.~\onlinecite{Esslinger_strong_int} is the result of a subtle interface effect at the ends of the ballistic wire, that becomes superconducting due to confinement-induced renormalization of interactions. Since SC penetrates in the quasi-two dimensional part of the lead, channels that would otherwise be reflected by the constriction can \martonnew{participate in Andreev processes, thus delivering Cooper pairs to the condensate which propagate through the interior part of the channel as a spinless superfluid.} We could also explain non-monotonicities in the spin-conductance curve as the gate potential was changed, and predict additional non-monotonicities of the conductance as a function of the confinement frequency at low temperatures.

\begin{acknowledgments}
\emph {Acknowledgements:}
Enlightening discussions with  F. Pientka, S. Gopalakrishnan, J.-P. Brantut, M. Lebrat, S. Krinner and D. Husmann are gratefully acknowledged.
The work of E. D. and M. K.-N. was supported by the Harvard-MIT CUA, NSF Grant No. DMR-1308435,  AFOSR Quantum Simulation MURI, the ARO-MURI on Atomtronics, and ARO MURI Quism program. 
E. D. also acknowledges support from Dr.~Max R\"ossler, the Walter Haefner Foundation, the ETH Foundation, the Simons Foundation and the Humboldt Foundation.
L. G. was supported by the DOE contract DE-FG02-08ER46482. 
T.E. acknowledges the Staatssekretariat f\"ur Bildung, Forschung und Innovation SBFI for the support of the Horizon2020 project Quantum simulations of insulators and conductors QUIC.
\end{acknowledgments}

\end{document}